\documentclass[a4paper,11pt]{article}
\usepackage{amsmath,amssymb,indentfirst,epsfig,multirow,amsthm,color,array,amsfonts,graphicx,fullpage,hyperref,bm}
\usepackage{subfigure,upgreek, booktabs}
\usepackage[latin1]{inputenc}
\newcommand{\ssection}[1]{\section[#1]{\centering #1}}
\newcommand{\ssubsection}[1]{\subsection[#1]{\centering #1}}

\begin{document}
\title{{\bf The Kumaraswamy Pareto distribution}}
\author{Marcelo Bourguignon$^{a,}$\footnote{Corresponding author. E-mail: m.p.bourguignon@gmail.com}, Rodrigo B. Silva$^{a,}$\footnote{E-mail: rodrigobs29@gmail.com}, Luz M. Zea$^{a,}$\footnote{E-mail: milezea@gmail.com}\,\, and Gauss M. Cordeiro$^{a,}$\footnote{E-mail: gauss@de.ufpe.br}\\\\
\centerline{\small{\it $^a$Departamento de Estat\'\i stica,
Universidade Federal de Pernambuco,
}}\\
\centerline{\small{\it Cidade Universit\' aria - 50740-540, Pernambuco-PE, Brasil.}}\\
}
\date{}
\sloppy
\maketitle

\begin{abstract}
The modeling and analysis of lifetimes is an important aspect of
statistical work in a wide variety of scientific and technological
fields. For the first time, the called Kumaraswamy Pareto
distribution, is introduced and studied. The new distribution can
have a decreasing and upside-down bathtub failure rate function
depending on the values of its parameters. It in\-clu\-des as
special sub-models the Pareto and exponentiated Pareto (Gupta {\it et al}.~\cite{guptaetal1998}). distributions. Some structural properties of the
proposed distribution are studied in\-clu\-ding explicit expressions
for the moments and generating function. We provide the density
function of the order statistics and obtain their moments. The
method of maximum likelihood is used for estimating the model
parameters and the observed information matrix is derived. A real
data set is used to compare the new model with widely known
distributions.
\end{abstract}

\ssection{Introduction}
The Pareto
distribution is a very popular model named after a professor of
economics: Vilfredo Pareto. The various forms of the Pareto
distribution are very versatile and a variety of uncertainties can
be usefully modelled by them. For instance, they arise as tractable
`lifetime' models in actuarial sciences, economics, finance, life
testing and climatology, where it usually describes the occurrence
of extreme weather.

The random variable $X$ has the Pareto distribution if its cumulative distribution function (cdf) for $x\geq \beta$ is
given by
\begin{equation}\label{CDFEP}
G(x;\beta,k) = 1-\left(\frac{\beta}{x}\right)^k,
\end{equation}
where $ \beta >0$ is a scale parameter and $k > 0$ is a shape
parameter. The probability density function (pdf) corresponding to
(\ref{CDFEP}) is

\begin{equation*}\label{PDFEP}
g(x;\beta,k) = \frac{k\,\beta^k}{x^{k+1}}.
\end{equation*}

Several generalized forms of the Pareto distribution can be found in
the literature. The term ``ge\-ne\-ra\-li\-zed Pareto'' (GP) distribution
was first used by Pickands~\cite{pickands1975} when making statistical inferences
about the upper tail of a distribution function. As expected, the
Pareto distribution can be seen as a special case of the GP
distribution. It can also be obtained as a special case of another
extended form generated by compounding a heavy-tailed skewed
conditional gamma density function with parameters $\alpha$ and
$\beta^{-1}$, where the weighting function for $\beta$ has a gamma
distribution with parameters $k$ and $\theta$ (Hogg {\it et al.}~\cite{hoggetal2005}).

Gupta {\it et al}.~\cite{guptaetal1998} extended the Pareto distribution by
raising (\ref{CDFEP}) to a positive power. In this note, we refer to
this extension as the exponentiated Pareto (EP) distribution.
Recently, many authors have considered various exponentiated-type
distributions based on some known distributions such as the
exponential, Rayleigh, Weibull, gamma and Burr distributions; see,
for example, Gupta and Kundu (\cite{guptakundu1999}, \cite{guptakundu2001a} and \cite{guptakundu2001b}), Surles and Padgett~\cite{surlespadgett2001}, Kundu and Raqab~\cite{kunduraqab2005} and Silva {\it et al.}~\cite{silvaetal2010}. The methods of moments and maximum likelihood have been used to fit
these models.

Further, Akinsete \emph{et al.}~\cite{akinseteetal2008} and Mahmoudi ~\cite{mahmoudi2011} extended the Pareto and GP distributions by defining the beta Pareto (BP) and
beta generalized Pareto (BGP) distributions, respectively, based on the class of generalized (so-called ``beta-G")   distributions introduced by Eugene {\it et al.}~\cite{eugeneetal2002}. The generalized distributions are obtained by taking any parent G distribution in the cdf of a beta distribution with two additional shape parameters, whose role is to introduce skewness and to vary tail weight. Following the same idea, many beta-type distributions were introduced and studied, see, for example, Barreto-Souza {\it et al.}~\cite{barretosouzaetal2010}, Silva {\it et al.}~\cite{silvaortegacordeiro2010} and Cordeiro {\it et al.}~\cite{cordeiroetal2011}.

In this context, we propose an extension of the Pareto distribution based on the fa\-mi\-ly of Kumaraswamy generalized (denoted with the prefix ``Kw-G" for short) distributions introduced by Cordeiro and de Castro~\cite{cordeirocastro2011}. Nadarajah {\it et al.}~\cite{nadarajahetal2011} studied some mathematical properties of this family. The Kumaraswamy (Kw) distribution is not very common among statisticians and has been little explored in the literature. Its cdf (for $0<x<1$) is $F(x) = 1 - (1-x^a)^b$, where $a>0$ and $b>0$ are shape parameters, and the density function has a simple form $f(x)= a\,b\,x^{a-1}(1-x^a)^{b-1}$, which can be unimodal, increasing, decreasing or constant, depending on the parameter values. It does not seem to be very familiar to statisticians and has not been investigated systematically in much detail before, nor has its relative interchangeability with the beta distribution been widely appreciated. However, in a very recent paper, Jones~\cite{jones2009} explored the background 
and genesis of this distribution and, more importantly, made clear some similarities and differences between the beta and Kw distributions.

In this note, we combine the works of Kumaraswamy~\cite{kumaraswamy1980} and Cordeiro and de Castro~\cite{cordeirocastro2011} to derive some mathematical properties of a new model, called the Kumaraswamy Pareto (Kw-P) distribution, which stems from the following general construction: if G denotes the baseline cumulative function of a random variable, then a generalized class of distributions can be defined by
\begin{equation}\label{cdfkuma}
F(x) = 1 - \left[1 - G(x)^a\right]^b,
\end{equation}
where $a > 0$ and $b > 0$ are two additional shape parameters which
govern skewness and tail weights. Because of its tractable
distribution function (\ref{cdfkuma}), the Kw-G distribution can be
used quite effectively even if the data are censored. Correspondingly, its density function is
distributions has a very simple form
\begin{equation}\label{pdfkuma}
f(x) = a\,b\,g(x)\,G(x)^{a-1}\left[1 - G(x)^a\right]^{b-1}.
\end{equation}

The density family (\ref{pdfkuma}) has many of the same properties
of the class of beta-G distributions (see Eugene {\it et al.}~\cite{eugeneetal2002}), but has some advantages in terms of tractability, since it
does not involve any special function such as the beta function.

Equivalently, as occurs with the beta-G family of distributions, special Kw-G distributions can be generated as follows: the Kw-normal distribution is obtained by taking $G(x)$ in
(\ref{cdfkuma}) to be the normal cumulative function. Analogously, the Kw-Weibull (Cordeiro {\it et al.}~\cite{cordeiroorteganadarajah2010}), Kw-generalized gamma (Pascoa {\it et al.}~\cite{pascoaetal2011}), Kw-Birnbaum-Saunders (Saulo {\it et al.}~\cite{sauloleaobourguignon2012}) and Kw-Gumbel (Cordeiro {\it et al.}~\cite{cordeiroappear2011}) distributions are obtained by taking $G(x)$ to be the cdf of the Weibull, generalized gamma, Birnbaum-Saunders and Gumbel distributions, respectively, among several others. Hence, each new Kw-G distribution can be generated from a specified G distribution.

This paper is outlined as follows. In section 2, we define the Kw-P distribution and provide expansions for its cumulative and density functions. In addition, we study the limit behavior of its pdf and hazard rate function. A range of mathematical properties of this distribution is considered in sections 3-7. These include quantile function, simulation, skewness and kurtosis, order  statistics, generating and characteristic functions, incomplete moments, L-moments and mean deviations. The Rényi entropy is calculated in section 8. Maximum likelihood estimation is performed and the observed information matrix is determined in section 9. In section 10, we provide an application of the Kw-P distribution to a flood data set. Finally, some conclusions are addressed in section 11.

\ssection{The Kw-P distribution}
If $G(x;\beta,k)$ is the Pareto cumulative distribution with parameters $\beta$ and $k$, then equation (\ref{cdfkuma}) yields the Kw-P cumulative distribution (for $x \geq \beta$)
\begin{equation}\label{eq2}
F(x;\beta,k,a,b) = 1 - \left\{1 -
\left[1-\left(\frac{\beta}{x}\right)^{k}\right]^{a} \right\}^{b},
\end{equation}
where $\beta>0$ is a scale parameter and the other positive
parameters $k,a$ and $b$ are shape parameters. The corresponding pdf
and hazard rate function are
\begin{equation}\label{fdpkwep}
f(x;\beta,k,a,b) = \frac{a\,b\, k\,
\beta^k}{x^{k+1}}\left[1-\left(\frac{\beta}{x}\right)^k\right]^{a-1}\left\{1-\left[1-\left(\frac{\beta}{x}\right)^k
\right]^{a}\right\}^{b-1},
\end{equation}
and
\begin{equation}\label{hazardkwep}
\tau(x;\beta,k,a,b) = \frac{a\,b\, k\,
\beta^k\left[1-\left(\beta/x\right)^k\right]^{a-1}}{x^{k+1}\,\{1 -
[1-(\beta/x)^{k} ]^{a} \}},
\end{equation}
respectively.

The Kw-P distribution is not in fact very tractable. However, its heavy tail can adjust skewed data that cannot be properly fitted by existing distributions. Furthermore, the cumulative
and hazard rate functions are simple.

\begin{figure}[!htbp]
\centering
   \begin{tabular}{cc}
    \includegraphics[width=0.47\textwidth]{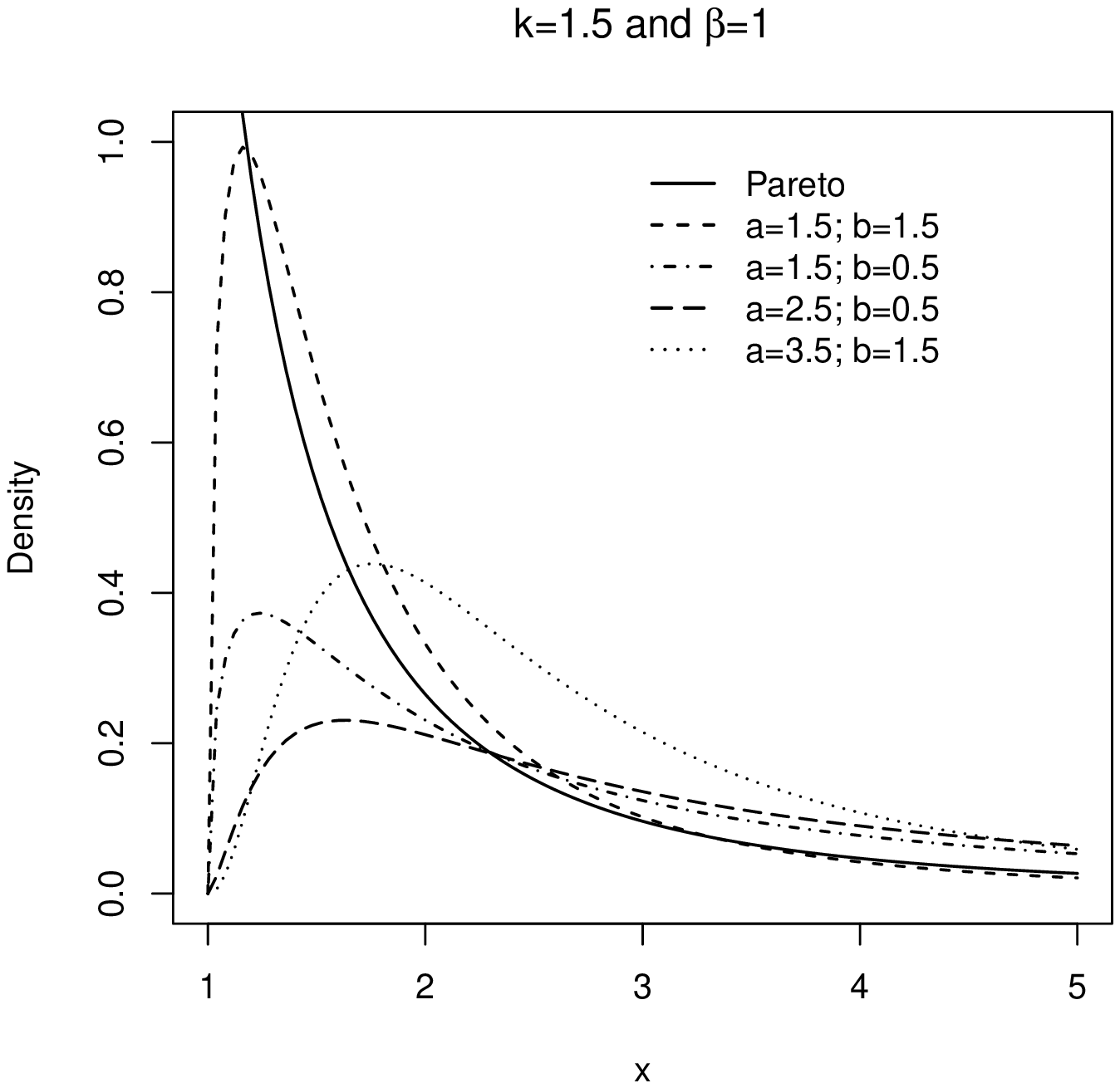}&\includegraphics[width=0.47\textwidth]{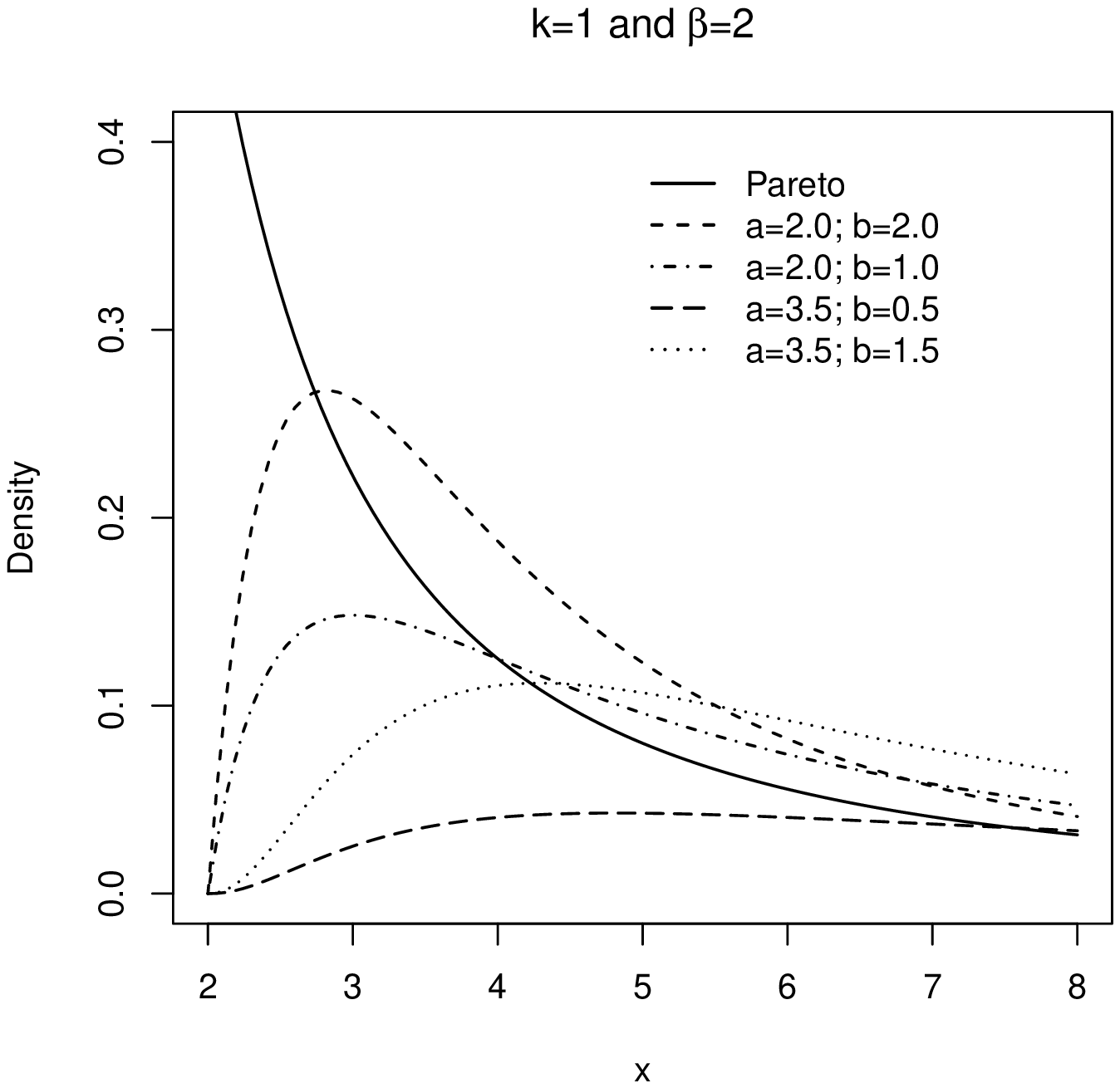}\\
     \end{tabular}
        \caption{Plots of the Kw-P density function for some parameter values.}
    \label{fig:kwpdensity}
\end{figure}

\begin{figure}[!htbp]
\centering
   \begin{tabular}{cc}
     \includegraphics[width=0.47\textwidth]{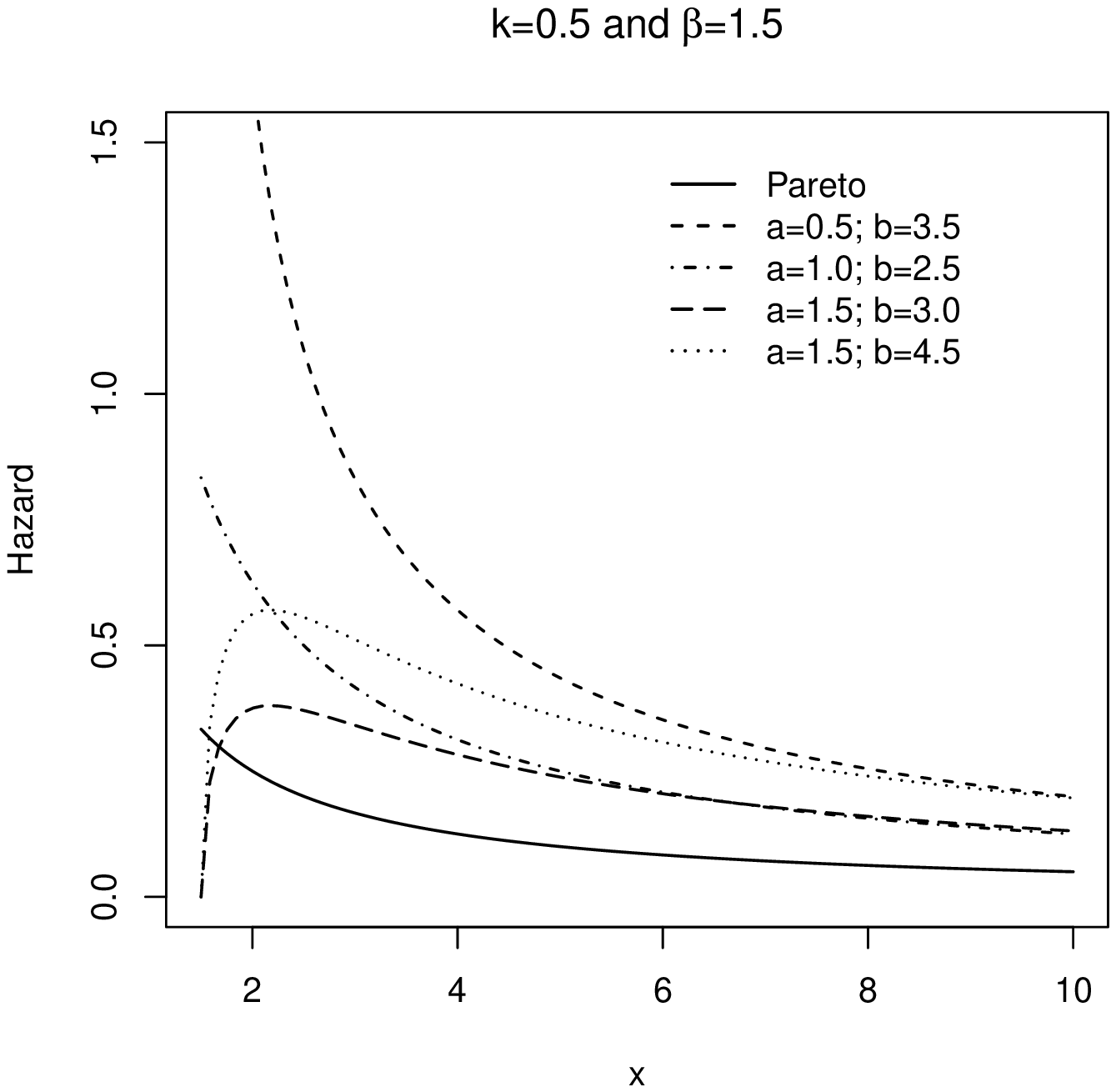}&\includegraphics[width=0.47\textwidth]{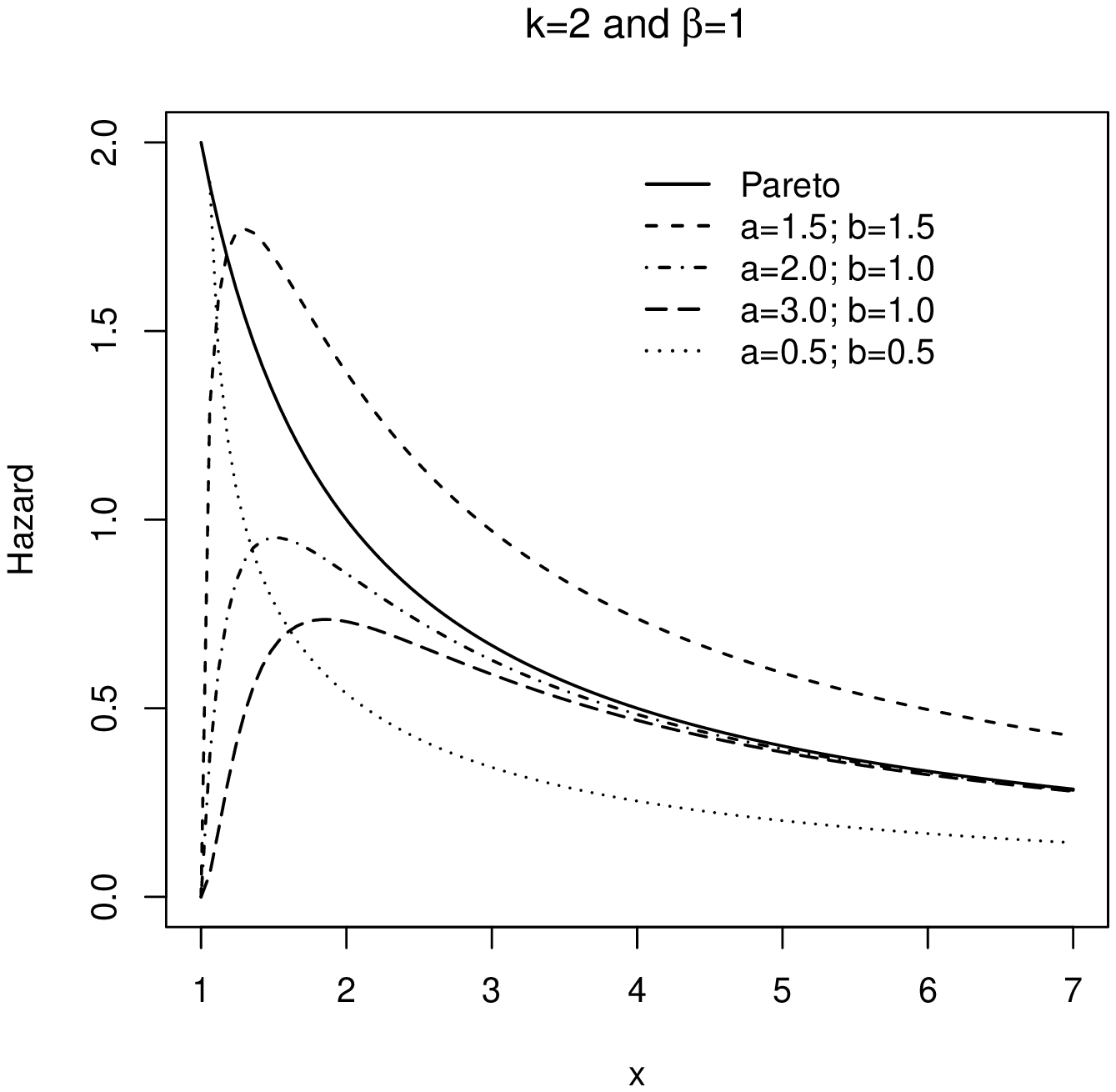}\\
   \end{tabular}
        \caption{Plots of the Kw-P hazard function for some parameter values.}
    \label{fig:kwphazard}
\end{figure}

In Figures \ref{fig:kwpdensity} and \ref{fig:kwphazard}, we plot the density and failure rate functions of the Kw-P distribution for selected parameter values, respectively. We can verify that this distribution can have a decreasing and upside-down bathtub failure rate function depending on the values of its pa\-ra\-me\-ters.

\ssubsection{Expansions for the cumulative and density functions}

Here, we give simple expansions for the Kw-P cumulative distribution. By using the ge\-ne\-ra\-li\-zed binomial theorem (for $0 < a < 1$)
\begin{equation}\label{bin}
(1+a)^\nu = \sum_{i=0}^{\infty} \binom{\nu}{i}\,a^i,
\end{equation}
where $$\binom{\nu}{i}=\frac{n(n-1)\ldots(\nu-i+1)}{i!},$$ in
equation (\ref{eq2}), we can write
\begin{eqnarray*}\label{cdfkwep} \nonumber
F(x;\beta,k,a,b) &=& 1 -
\sum^{\infty}_{i=0}(-1)^i\,\binom{b}{i}\left[1-\left(\frac{\beta}{x}\right)^k
\right]^{ai}\\
&=& 1 - \sum^{\infty}_{i=0}\eta_i\,H(x;\beta,k,ia),
\end{eqnarray*}
where $\eta_i = (-1)^i\binom{b}{i}$ and $H(x;\beta, k, ia)$ denotes the EP cumulative distribution (with pa\-ra\-me\-ters $\beta$, $k$ and $ia$) given by
\[H(x;\beta,k,\alpha)=\left[ 1-\left(\frac{\beta}{x}\right)^k\right]^\alpha.\]
Now, using the power series (\ref{bin}) in the last term of
(\ref{fdpkwep}), we obtain
\begin{eqnarray}\label{lcpareto}\nonumber
f(x;\beta,k,a,b) &=& \frac{a\, b\, k\,\beta^k}{x^{k+1}}\sum^{\infty}_{i=0}(-1)^i\,\binom{b-1}{i}\,\left[1-\left(\frac{\beta}{x}\right)^k\right]^{ a(i+1)-1}\\
&=& \sum_{j=0}^{\infty}w_{j}\, g(x; \beta, k(j+1)),
\end{eqnarray}
where
\begin{equation}
 w_{j} = \frac{a\, b}{(j+1)}\sum_{i=0}^{\infty}(-1)^{i+j}\binom{b-1}{i}\binom{a(i+1)-1}{j} \nonumber
\end{equation}
and $g(x; \beta, k(j+1))$ denotes the Pareto density function with parameters $\beta$ and $k(j+1)$ and cumulative distribution as in (\ref{CDFEP}). Thus, the Kw-P density function can be expressed as an infinite linear combination of Pareto densities. Thus, some of its mathematical properties can be obtained directly from those properties of the Pareto distribution. For example, the ordinary, inverse and factorial moments, moment generating function (mgf) and characteristic function of the Kw-P distribution follow immediately from those quantities of the Pareto distribution.

\ssubsection{Limiting behaviour of Kw-P density and hazard functions}

\noindent \textbf{Lemma 1.} The limit of the Kw-P density function as $x\rightarrow \infty$ is 0 and the limit as $x\rightarrow \beta$ are
\begin{displaymath}
\lim_{x\rightarrow \beta}f(x;\beta,k,a,b) = \left\{
\begin{array}{ll}
\infty,  & \textrm{for} \, \, 0 < a < 1,\\
\dfrac{b k}{\beta}, & \textrm{for} \, \, a = 1,\\
0, & \textrm{for} \, \, a > 1.\\
\end{array} \right.
\end{displaymath}

\noindent \emph{Proof.} It is easy to demonstrate the result from the density function (\ref{fdpkwep}).\\

\noindent \textbf{Lemma 2.} The limit of the Kw-P hazard function as $x\rightarrow \infty$ is 0 and the limit as $x\rightarrow \beta$ are
\begin{displaymath}
\lim_{x\rightarrow \beta}\tau(x;\beta,k,a,b) = \left\{
\begin{array}{ll}
\infty,  & \textrm{for} \, \, 0 < a < 1,\\
\dfrac{b k}{\beta}, & \textrm{for} \, \, a = 1,\\
0, & \textrm{for} \, \, a > 1.\\
\end{array} \right.
\end{displaymath}

\noindent \emph{Proof.} It is straightforward to prove this result from (\ref{hazardkwep}).

\ssection{Moments and generating function}

Here and henceforth, let $X$ be a Kw-P random variable following (\ref{fdpkwep}).
 \vskip 3mm

\ssubsection{Moments}

The $r$th moment of $X$ can be obtained from (\ref{lcpareto}) as
\begin{eqnarray}\label{momentkwep} \nonumber
\mbox{E}(X^r)  &=&\sum^{\infty}_{j=0} w_j \int_{\beta}^{\infty}x^r \, g(x; \beta, k(j+1)) dx\\
&=&k\,\beta^r\sum^{\infty}_{j=0}\frac{(j+1)\,w_{j}}{\left[k(j+1)-r\right]},
\end{eqnarray}
for $ r < bk$. In particular, setting $r = 1$ in (\ref{momentkwep}), the mean of $X$ reduces to
\begin{equation}\label{meankwep}
\mu = \mbox{E}(X) = k\,\beta
\sum^{\infty}_{j=0}\frac{(j+1)\,w_{j}}{\left[k(j+1)-1\right]},
\quad\text{for}\ \  bk > 1.
\end{equation}

Setting $a=b=1$, we have
\begin{displaymath}
w_j = \left\{ \begin{array}{ll}
1, & \textrm{for $j=0$},\\
0, & \textrm{for $j \geq 1$\,.}\\
\end{array} \right.
\end{displaymath}
Then, equation (\ref{meankwep}) reduces to (for $k>1$)
\begin{equation}
\mbox{E}(X) = \frac{k\beta}{k-1}, \nonumber
\end{equation}
which is precisely the mean of the Pareto distribution.

\ssubsection{Incomplete moments}
 If $Y$ is a random variable with a Pareto distribution with parameters $\beta$ and $k$, the $r$th incomplete moment of $Y$, for $r< k$, is given by
\begin{eqnarray*}\label{inmoment}
M_r(z) = \int^{z}_{\beta}y^r\, g(y;\beta,k)dy  =
\frac{k\beta^r}{(k-r)}\,\Biggl[1-\biggl(\frac\beta
z\biggr)^{k-r}\Biggr]\,.
\end{eqnarray*}

From this equation, we note that $M_r(z) \rightarrow E(Y^r)$ when $z \rightarrow \infty$, whenever $k>r$. Let X $\sim$ Kw-P($\beta,k,a,b$). The $r$th incomplete moment of $X$ is then equal to
\begin{eqnarray}\label{momentincompkwep}
M_r(z) = \int^{z}_{\beta}x^r f(x;\beta,k,a,b)dx = k\,
\beta^r\sum^{\infty}_{j=0}\frac{(j+1)\,w_{j}}{[k(j+1)-r]}\,\Biggl[1-\biggl(\frac\beta
z\biggr)^{k(j+1)-r}\Biggr]\,,
\end{eqnarray}
which provided that $r< bk$.
\ssubsection{Generating function}
First, the mgf $M_Y(t)$ corresponding to a random variable $Y$ with Pareto distribution with parameters $\beta$ and $k$ is only defined for non-positive values of $t$. It is given by
\begin{eqnarray} \nonumber
M_Y(t)
=k\,(-\beta t)^k\,\Gamma(-k,-\beta t)\,,\, &\text{if} \,\, t<0,
\end{eqnarray}
where $\Gamma(\cdot,\cdot)$ denotes the incomplete gamma function
\[\Gamma(s,x)=\int_x^\infty t^{s-1}e^{-t}dt\,.\]

Thus, using $M_Y(t)$ and (\ref{lcpareto}), we can write for $t<0$
\begin{eqnarray}\label{mgfkwep} \nonumber
M_{X}(t)&=&  \sum_{j=0}^{\infty} w_j \int_{\beta}^{\infty}\mathrm{e}^{tx}\,g(x; \beta, k(j+1))dx \nonumber\\
&=&  k\sum_{j=0}^{\infty} (j+1)(-\beta t)^{k(j+1)}\,w_j\,\Gamma(-k(j+1),-\beta t).
\end{eqnarray}

In the same way, the characteristic function of the Kw-P distribution becomes $\phi_X(t)=M_X(it)$, where $i=\sqrt{-1}$ is the unit imaginary number.

\ssection{Quantile function and simulation}
We present a method for simulating from the Kw-P distribution (\ref{fdpkwep}). The quantile function corresponding to (\ref{eq2}) is
\begin{equation}\label{quantile}
Q(u) = F^{-1}(u) = \frac{\beta}{\{ 1 - [1 - (1 - u)^{1/b}]^{1/a}
\}^{1/k}}.
\end{equation}

Simulating the Kw-P random variable is straightforward. Let $U$ be a uniform variate on the unit interval $(0,1)$. Thus, by means of the inverse transformation method, we consider the random variable $X$ given by
\begin{equation*}
X = \frac{\beta}{\{ 1 - [1 - (1 - U)^{1/b}]^{1/a} \}^{1/k}},
\end{equation*}
which follows (\ref{fdpkwep}), i.e.,  $X \sim \operatorname{Kw-P}(\beta,k,a,b)$.

\begin{figure}[!htbp]
\centering
\subfigure[]{\includegraphics[scale=0.5]{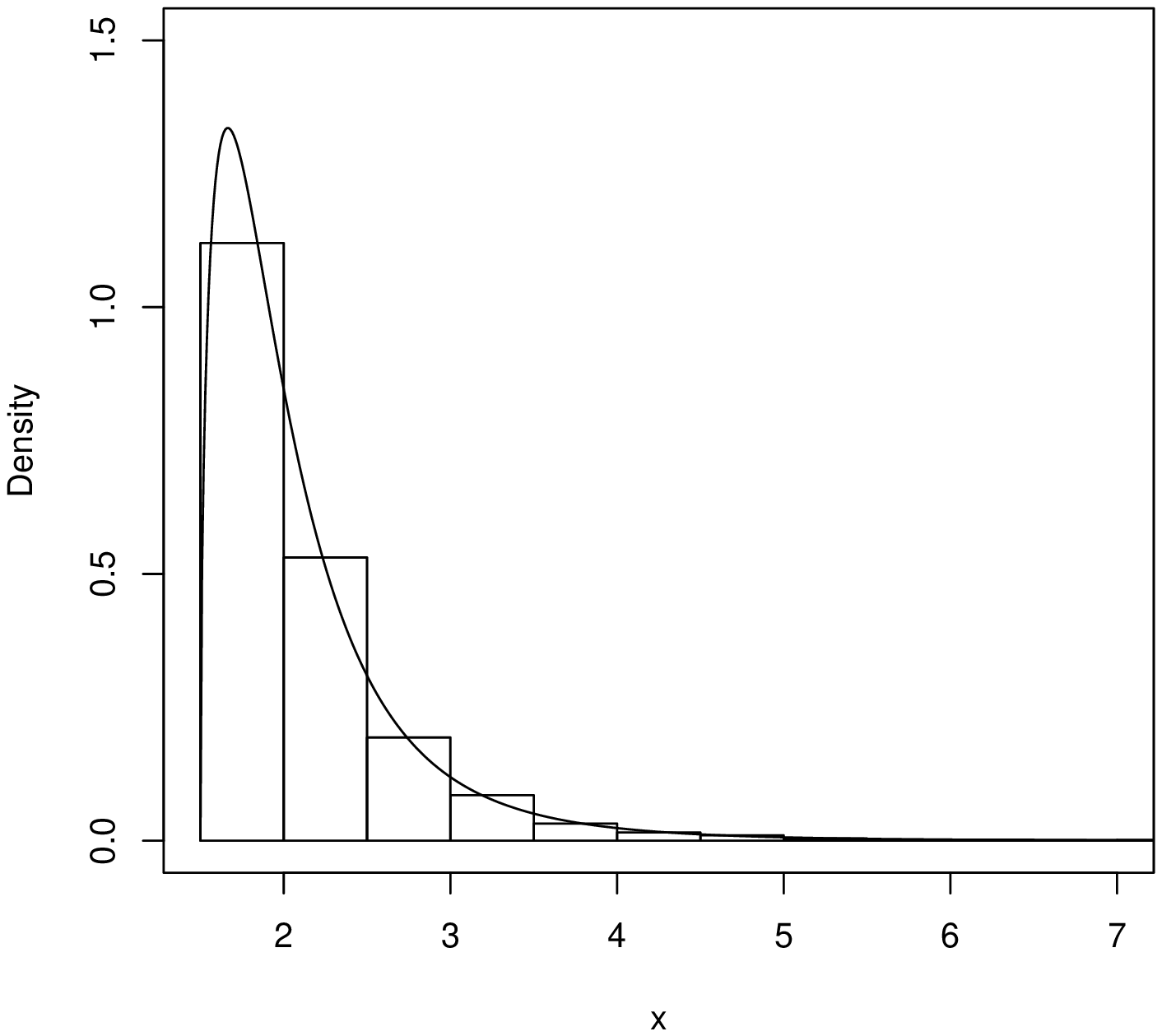}}
\subfigure[]{\includegraphics[scale=0.5]{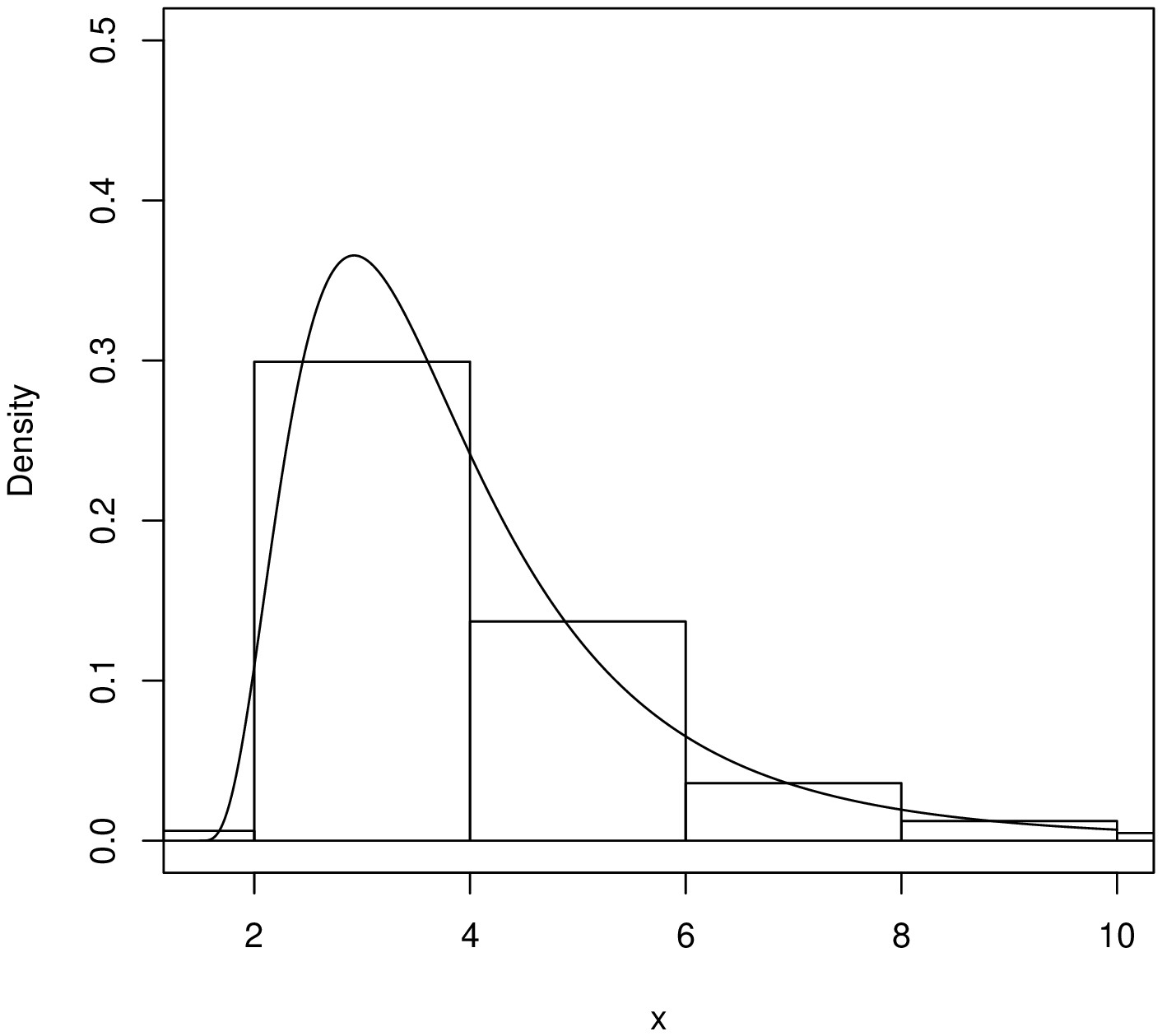}}
\caption{Plots of the Kw-P densities for simulated data sets: (a)
$a=1.5, b=3.5, \beta = 1.5, k = 1.5$ and (b) $a=5.0, b=3.0, \beta =
1.5, k = 1.5$.}
 \label{figapli}
\end{figure}

The plots comparing the exact Kw-P densities and histograms from two simulated data sets for some parameter values are given in Figure \ref{figapli}. These plots indicate that the simulated values are consistent with the Kw-P  theoretical density function.

\ssection{Skewness and Kurtosis}

The shortcomings of the classical kurtosis measure are well-known. There are many heavy-tailed distributions for which this measure is infinite. So, it becomes uninformative precisely when it needs to be. Indeed, our motivation to use quantile-based measures stemmed from the non-existance of classical kurtosis for many of the Kw distributions.

The Bowley's skewness (see Kenney and Keeping~\cite{kk1962}) is based on quartiles:
$$B=\frac{Q(3/4)-2Q(1/2)+Q(1/4)}{Q(3/4)-Q(1/4)}$$
and the Moors' kurtosis (see Moors~\cite{moors1998}) is based on octiles:
$$M=\frac{Q(7/8)-Q(5/8)-Q(3/8)+Q(1/8)}{Q(6/8)-Q(2/8)},$$
where $Q(\cdot)$ represents the quantile function.

\begin{figure}
\centering
\subfigure[]{\epsfig{file=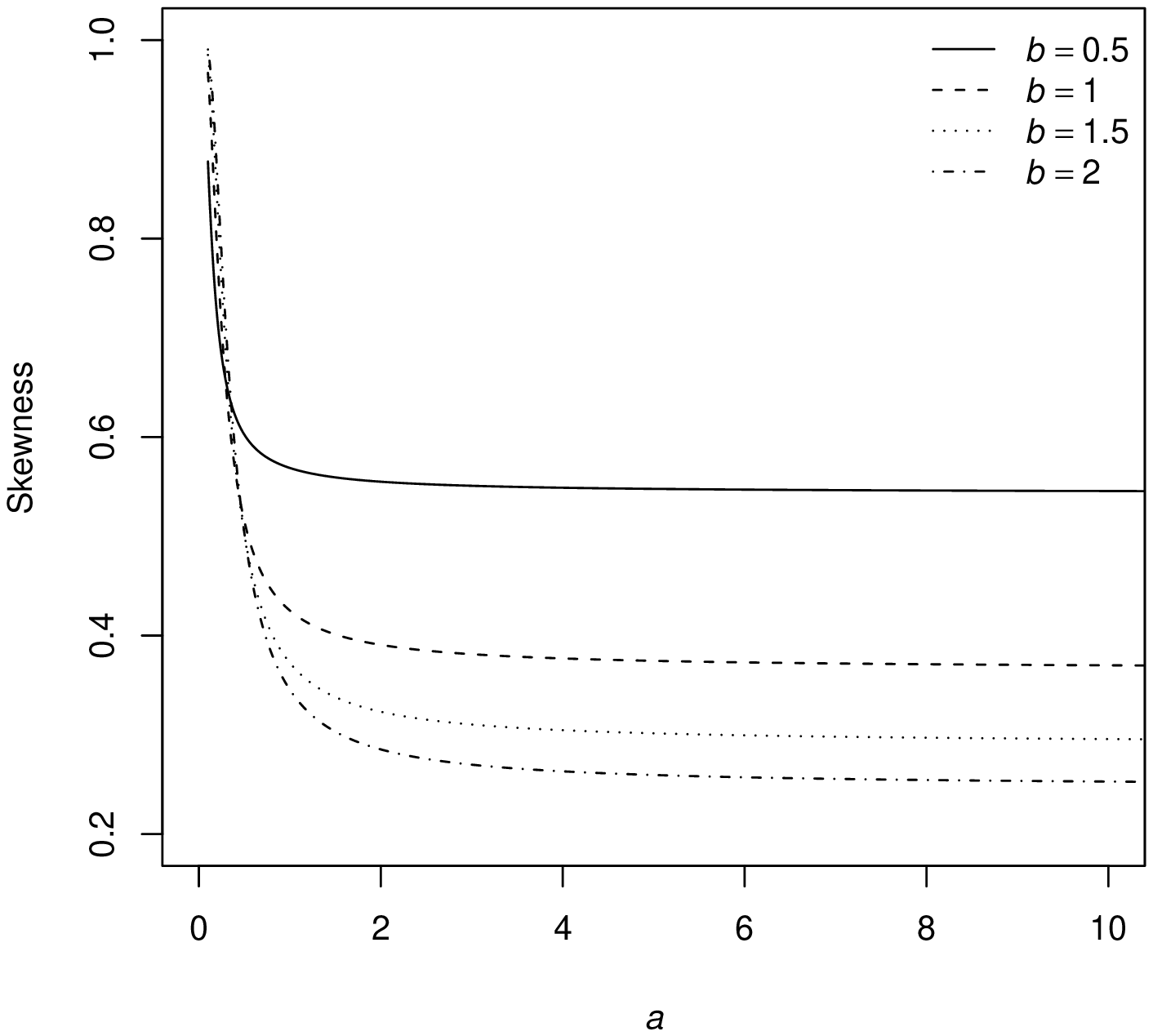,height=0.45\linewidth}}
\subfigure[]{\epsfig{file=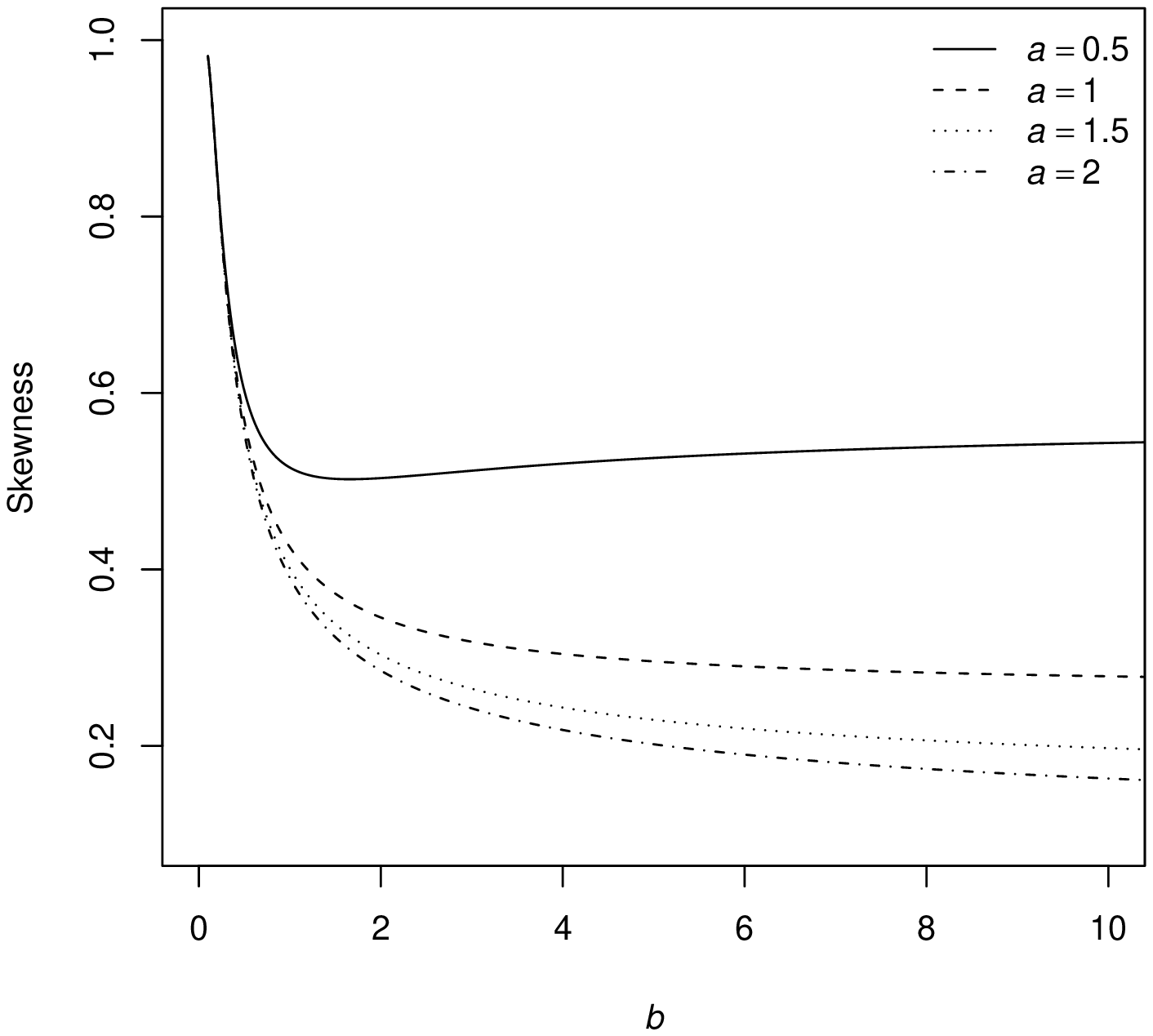,height=0.45\linewidth}}
\subfigure[]{\epsfig{file=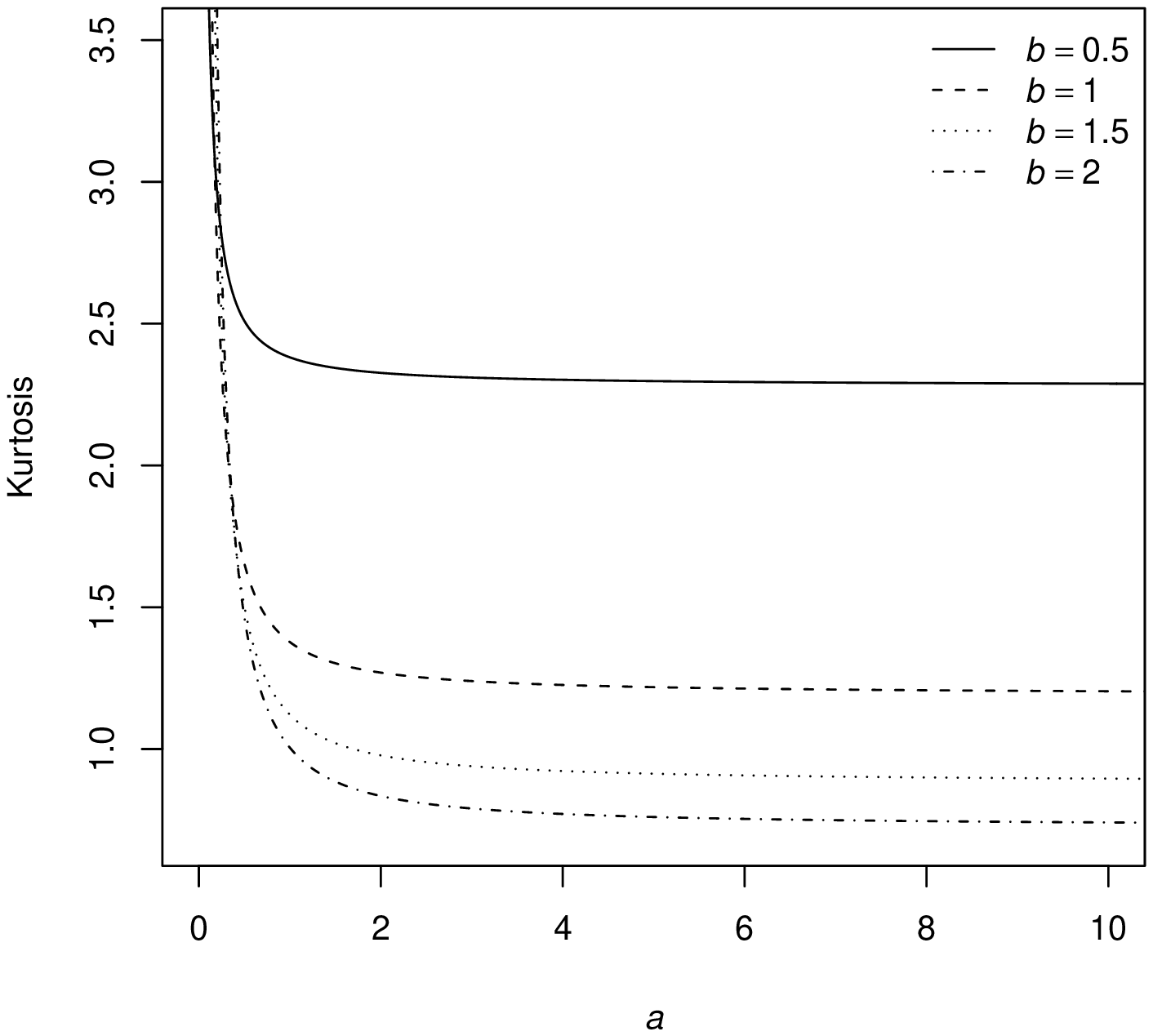,height=0.45\linewidth}}
\subfigure[]{\epsfig{file=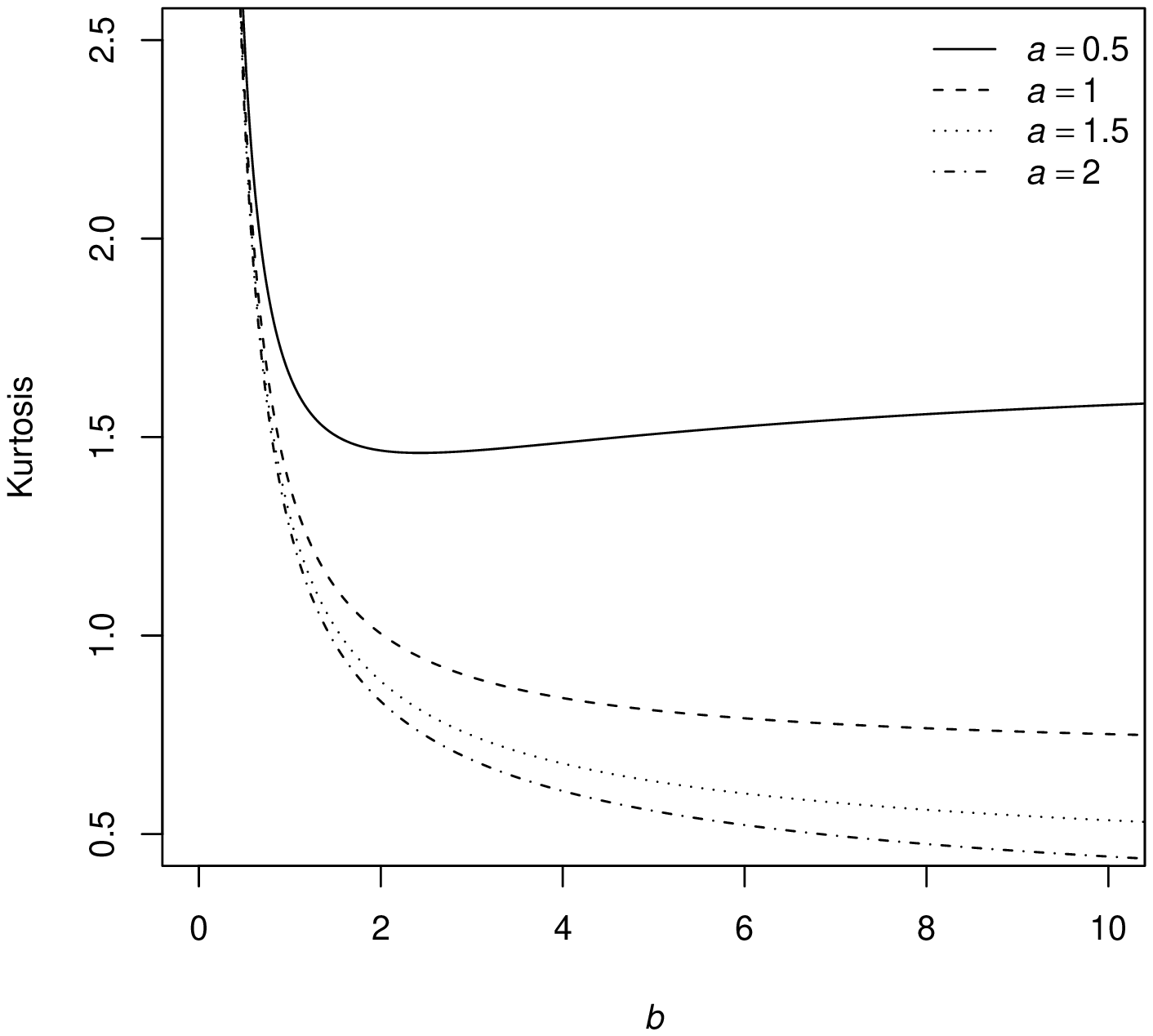,height=0.45\linewidth}}
\hfill\caption{Plots of the Kw-P skewness and kurtosis as a function
of $a$ for selected values of $b$ and as a function of $b$ for
selected values of $a$.} \label{ff4}
\end{figure}

Plots of the skewness and kurtosis for some choices of the parameter $b$ as function of $a$, and for some choices of the parameter $a$ as function of $b$, for $\beta = 1.0$ and $k=1.5$, are shown in Figure \ref{ff4}. These plots show that the skewness and kurtosis decrease when $b$ increases for fixed $a$ and when $a$ increases for fixed $b$.

\ssection{Order statistics}
Moments of order statistics play an important role in quality control testing and reliability, where a practitioner needs to predict the failure of future items based on the times of a few
early failures. These predictors are often based on moments of order statistics. We now derive an explicit expression for the density function of the $i$th order statistic $X_{i:n}$, say $f_{i:n}(x)$, in a random sample of size $n$ from the Kw-P distribution. We can write
\begin{equation}\label{orderstatistics}
f_{i:n}(x) =
\frac{n!}{(i-1)!(n-i)!}f(x)\,F^{i-1}(x)[1-F(x)]^{n-i},\nonumber
\end{equation}
where $f(\cdot)$ and $F(\cdot)$ are the pdf and cdf of the Kw-P distribution, respectively. From the above equation and using the series representation (\ref{bin}) repeatedly, we obtain a useful expression for $f_{i:n}(x)$ given by
\begin{equation}\label{ith order}
f_{i:n}(x) = \sum_{r=0}^{\infty}c_{i:n}^{(r)}\,\,g(x; k(r+1), \beta),
\end{equation}
where
\begin{equation}
c_{i:n}^{(r)} =
\frac{n!\,a\,b}{(i-1)!(n-i)!}\sum_{l=0}^{\infty}\sum_{m=0}^{\infty}
\frac{(-1)^{l+m+r}}{r+1} \binom{i-1}{l}\,
\binom{b(n+l+1-i)-1}{m}\,\binom{a(m+1)-1}{r} \nonumber
\end{equation}
and $g(x; k(r+1), \beta)$ denotes the Pareto density function with parameters $k(r+1)$ and $\beta$. So, the density function of the order statistics is simply an infinite linear combination of Pareto densities. The pdf of the $i$th order statistic from a random sample of the Pareto distribution comes by setting $a=b=1$ in (\ref{ith order}). Evidently, equation (\ref{ith order}) plays an important role in the derivation of the main properties of the Kw-P order statistics.

For example, the $s$th raw moment of $X_{i:n}$ can be expressed as
\begin{equation}\label{ithorder}
E(X^s_{i:n}) = k \,\beta^s
\sum_{r=0}^{\infty}\frac{(r+1)c_{i:n}^{(r)}}{k(r+1)-s}.
\end{equation}

The L-moments are analogous to the ordinary moments but can be estimated by linear combinations of order statistics. They are linear functions of expected order statistics defined by
$$\lambda_{m+1} = \frac{1}{m+1} \sum_{k=0}^{m}(-1)^k\binom{m}{k}\,\operatorname{E}(X_{m+1-k:m+1}),\,\,m=0,1,\ldots$$

The first four L-moments are: $\lambda_1=\mbox{E}(X_{1:1})$, $\lambda_2=\frac{1}{2}\mbox{E}(X_{2:2}-X_{1:2})$, $\lambda_3=\frac{1}{3}\mbox{E}(X_{3:3}-2X_{2:3}+X_{1:3})$ and
$\lambda_4=\frac{1}{4}\mbox{E}(X_{4:4}-3X_{3:4}+ 3X_{2:4}-X_{1:4})$.
The L-moments have the advantage that they exist whenever the mean of the distribution exists, even though some higher moments may not exist, and are relatively robust to the effects of outliers.

From equation (\ref{ithorder}) with $s=1$, we can easily obtain explicit expressions for the L-moments of $X$.

\ssection{Mean deviations}

The mean deviations about the mean and the median can be used as measures of spread in a population. Let $\mu = E(X)$ and $m$ be the mean and the median of the Kw-P distribution, respectively. The mean deviations about the mean and about the median can be calculated as
\begin{equation}
 D(\mu) = E(|X-\mu|) = \int_{\beta}^{\infty}|x-\mu|\,f(x)dx \nonumber
\end{equation}
and
\begin{equation}
 D(m) = E(|X-m|) = \int_{\beta}^{\infty}|x-m|f(x)\,dx\,, \nonumber
\end{equation}
respectively. We obtain
\begin{eqnarray*}
 D(\mu) = \int_{\beta}^{\infty}|x-\mu|\,f(x)dx = 2 \mu F(\mu) - 2M_1(\mu)\,,
\end{eqnarray*}
where $M_1(\mu)$ denotes the first incomplete moment calculated from (\ref{momentincompkwep}) for $r=1$. Similarly, the mean deviation about the median follows as
\begin{eqnarray*}
 D(m) = \int_{\beta}^{\infty}|x-m|f(x)dx = \mu -2M_1(m)\,.
\end{eqnarray*}

\ssection{Rényi entropy}

The entropy of a random variable $X$ is a measure of uncertainty variation. The Rényi entropy is defined as
\begin{equation}
I_{R}(\delta) = \frac{1}{1-\delta}\log\left[I(\delta)\right],
\nonumber
\end{equation}
where $I(\delta)=\int_{\mathbb{R}}{f^{\delta}(x)dx}$, $\delta>0$ and
$\delta\neq1$. We have
\begin{equation}
I(\delta)= a^\delta b^\delta k^\delta
\beta^{k\delta}\int_{\beta}^{\infty}\frac{1}{x^{\delta(k+1)}}\left[1-\left(\frac{\beta}{x}\right)^k\right]^{\delta(a-1)}\left\{1-\left[1-\left(\frac{\beta}{x}\right)^k\right]^{a}\right\}^{\delta(b-1)}dx.
\nonumber
\end{equation}

By expanding the last term of the above integrand as in equation (\ref{bin}), we obtain
\begin{equation}
I(\delta)= a^\delta b^\delta k^\delta
\beta^{k\delta}\sum_{j=0}^{\infty}(-1)^{j}\binom{\delta(b-1)}{j}\int_{\beta}^{\infty}\frac{1}{x^{\delta(k+1)}}
\left[1-\left(\frac{\beta}{x}\right)^k\right]^{a(\delta+j)-\delta}dx.
\nonumber
\end{equation}

Transforming variables,  this equation becomes
\begin{equation}
I(\delta)= a^\delta b^\delta k^{\delta-1} \beta^{\delta+1}
\sum_{j=0}^{\infty}(-1)^{j}\,\binom{\delta(b-1)}{j}B\left(a(\delta+j)-\delta+1,
\frac{\delta(k+1)-1}{k}\right), \nonumber
\end{equation}
where $$B(a,b) = \int^{1}_{0}t^{a-1} (1-t)^{b-1}dt$$ denotes the beta function. Hence, the Rényi entropy reduces to
\begin{eqnarray*}
I_{R}(\delta) &=& \frac{\delta \log(a\,b)}{1-\delta} - \log k + \log \beta \\
&+& \frac{1}{1-\delta}\log
\sum_{j=0}^{\infty}(-1)^{j}\binom{\delta(b-1)}{j}B\left(a(\delta+j)-\delta+1,
\frac{\delta(k+1)-1}{k}\right).
\end{eqnarray*}

\ssection{Estimation and information matrix}

In this section, we discuss maximum likelihood estimation and
inference for the Kw-P distribution. Let $x_1, \ldots, x_n$ be a
random sample from $X \sim \mbox{Kw-P}(\beta, k, a, b)$ and let
$\boldsymbol{\theta} =(\beta, k, a, b)^\top$ be the vector of the
model parameters. The log-likelihood function for
$\boldsymbol{\theta}$ reduces to
\begin{eqnarray}\label{loglikelihood} \nonumber
\ell(\boldsymbol{\theta}) &=& n \log a + n \log b  + n\log k + nk\log \beta -(k+1)\sum_{i=1}^{n}\log(x_i)\\
 &+& (a-1)\sum_{i=1}^{n} \log \left[1-\left(\frac{\beta}{x_i}\right)^k\right] + (b-1)\sum_{i=1}^{n} \log \left\{1-
 \left[1-\left(\frac{\beta}{x_i}\right)^k\right]^{a}\right\}.
\end{eqnarray}

The score vector is $U(\boldsymbol{\theta})=(\partial\ell/\partial
k, \partial\ell/\partial a, \partial\ell/\partial b)^\top$, where
the components cor\-res\-pon\-ding to the model parameters are
calculated by dif\-fe\-ren\-tia\-ting (\ref{loglikelihood}). By
setting $z_i = 1- \left(\beta/x_i\right)^k$, we obtain

\begin{align*}\label{emva}
\frac{\partial \ell}{\partial k} &= \frac{n}{k} +
\frac{1}{k}\sum_{i=1}^{n}\log(1-z_i)  - \frac{(a-1)}{k}
\sum_{i=1}^{n} \frac{(1-z_i)\log(1-z_i)}{z_i} \notag\\
 &+ \frac{a (b-1)}{k} \sum_{i=1}^{n} \frac{z_i^{a-1}(1-z_i)\log(1-z_i)}{ (1-z_{i}^{a})}, \\
\frac{\partial \ell}{\partial a} &= \frac{n}{a} + \sum_{i=1}^{n}\log z_i - (b-1)\sum_{i=1}^{n}\frac{z_i^{a} \log z_i}{1 - z_i^{a}}\\
\intertext{and}
\frac{\partial \ell}{\partial b} &= \frac{n}{b} + \sum_{i=1}^{n}\log (1 - z_i^{a}). \\
\end{align*}

The maximum likelihood estimates (MLEs) of the parameters are the
solutions of the nonlinear equations $\nabla \ell=0$,  which are
solved iteratively. The observed information matrix given by
$$\boldsymbol{J}_n(\boldsymbol{\theta})=n\left[
\begin{array}{ccc}
J_{k k }&J_{k a}& J_{k b}\\
J_{a k }&J_{a a}& J_{a b}\\
J_{b k }&J_{b a}& J_{b b}\\
\end{array}\right],
$$
whose elements are
\begin{align*}
J_{kk} &= -\frac{n}{k^2} - \frac{2(a-1)}{k}\sum_{i=1}^{n}\frac{(1-z_i)\log(1-z_i)}{z_i^2} \\
&+ \frac{2a(b-1)}{k}\sum_{i=1}^{n}\frac{z_i^{a-1}(1-z_i)\log(1-z_i)\left[a-(a-1)(z_i^a+z_i^{-1})+(a-2)z_i^{a-1}\right]}{(1-z_i^a)^2},\\
J_{ka} &= \frac{(b-1)}{k}\sum_{i=1}^{n}\frac{z_{i}^{a
-1}(1-z_i)\log(1-z_i)\left[1-z_i^a - a\log
z_{i}\right]}{(1-z_{i}^{a})^2}
- \frac{1}{k} \sum_{i=1}^{n}\frac{(1-z_i)\log(1-z_i)}{z_i},\\
J_{kb} &= \frac{a}{k} \sum_{i=1}^{n}\frac{z_{i}^{a-1}(1-z_i)\log(1-z_i)}{1-z_{i}^{a}}, \quad J_{a a} = -\frac{n}{a^2} - 2(b-1) \sum_{i=1}^{n} \frac{z_i^{a}\log z_i}{(1-z_i^{a})^2}, \\
J_{a b} &= -\sum_{i=1}^{n} \frac{z_i^{a}\log z_i}{1-z_i^{a}} \quad \mbox{and} \quad J_{bb} = -\frac{n}{b^2}.\\
\end{align*}

\ssection{Simulation study and application}

In this section, we illustrate the usefulness of the Kw-P distribution.

\ssubsection{Simulation study}

We conduct Monte Carlo simulation studies to assess on the finite sample behavior of the MLEs of $\beta,k, a$ and $b$. All results were obtained from 1000 Monte Carlo replications and the simulations were carried out using the statistical software package R. In each replication, a random sample of size n is drawn from the Kw-P($\beta,k,a,b$) distribution and the BFGS method has been used by the authors for maximizing the total log-likelihood function $\ell(\boldsymbol{\theta})$. The Kw-P random number generation was performed using the inversion method. The true parameter values used in the data generating processes are $\beta = 1.5, k = 1.0, a = 0.5$ and $b = 2.5$. Table \ref{tabsimu} lists the means of the MLEs of the four parameters that index the Kw-P distribution along with the respective biases for sample sizes $n = 30, n = 50$ and $n = 100$.

The figures in Table \ref{tabsimu} indicate that the biases of the MLEs of $\beta, k, a,$ and $b$ decay toward zero as the sample size increases, as expected.

\ssubsection{The Wheaton River data}

The data correspond to the exceedances of flood peaks (in $\mbox{m}^3/$s) of the Wheaton River near Carcross in Yukon Territory, Canada. The data consist of 72 exceedances for the years 1958--1984, rounded to one decimal place. They were analysed by Choulakian and Stephens~\cite{choulakianstephens2001} and are listed in Table \ref{tab1}.
The distribution is highly skewed to the left. Recently, Akinsete {\it et al.}~\cite{akinseteetal2008} and Mahmoudi~\cite{mahmoudi2011} analysed these data using the BP and BGP distributions, respectively. We fit the Kw-P distribution to these data and compare the results with those by fitting some of its sub-models such as the EP and Pareto distributions, as well as the non-nested BP distribution. The required numerical evaluations are implemented using the SAS (PROCNLMIXED) and R softwares.

Tables \ref{tab2} and \ref{tab3} provide some descriptive statistics and the MLEs (with corresponding standard errors in parentheses) of the model parameters. Since $x \geq \beta$, the MLE of $\beta$ is the first-order statistic $x_{(1)}$, accordingly to Akinsete {\it et al.}~\cite{akinseteetal2008}. Since the values of the Akaike information criterion (AIC), Bayesian information criterion (BIC) and consistent Akaike information criterion (CAIC) are smaller for the Kw-P distribution compared with those values of the other models, the new distribution seems to be a very competitive model to these data.

\begin{figure}[!htbp]
\centering
\subfigure[]{\includegraphics[scale=0.5]{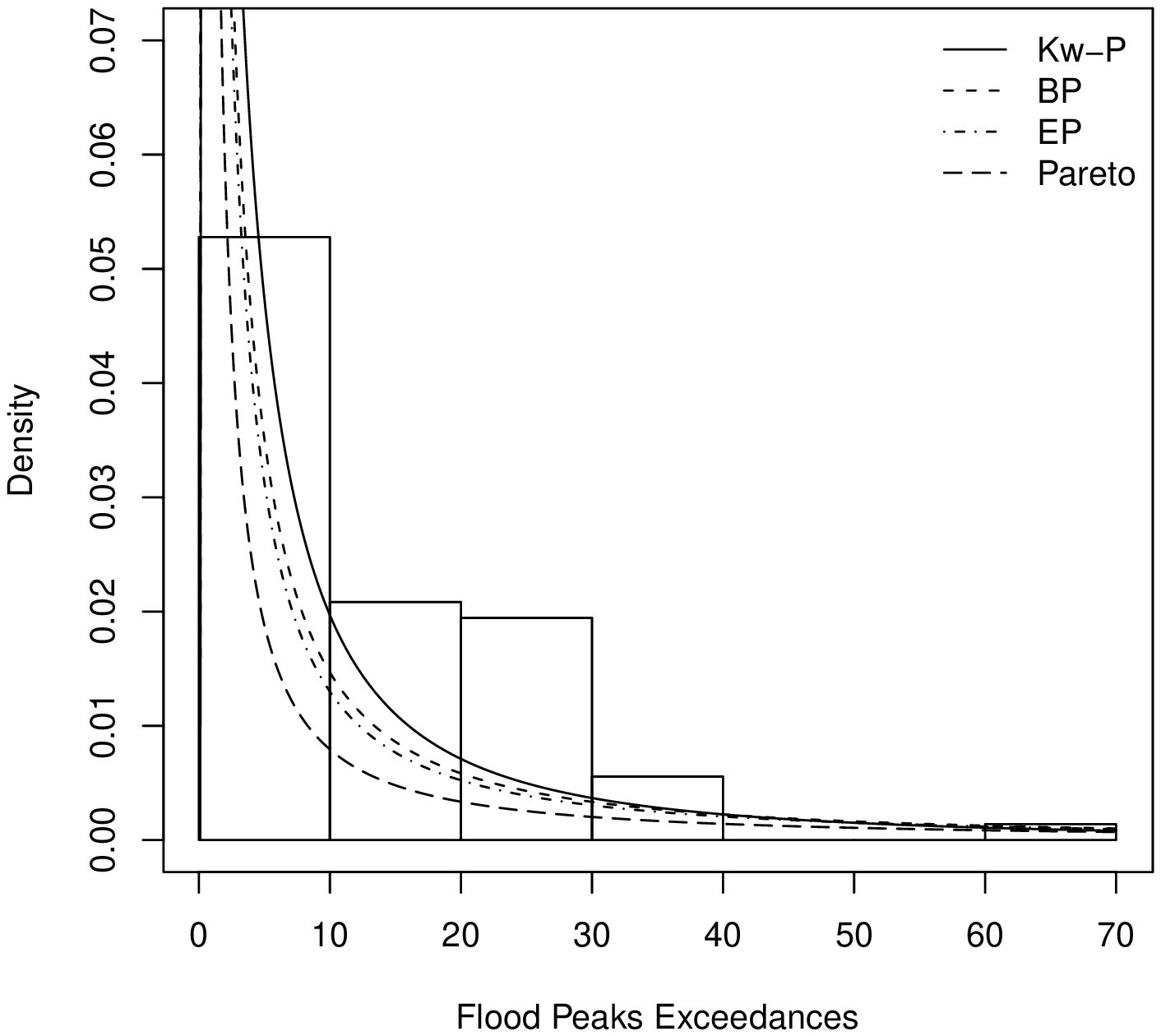}}
\subfigure[]{\includegraphics[scale=0.5]{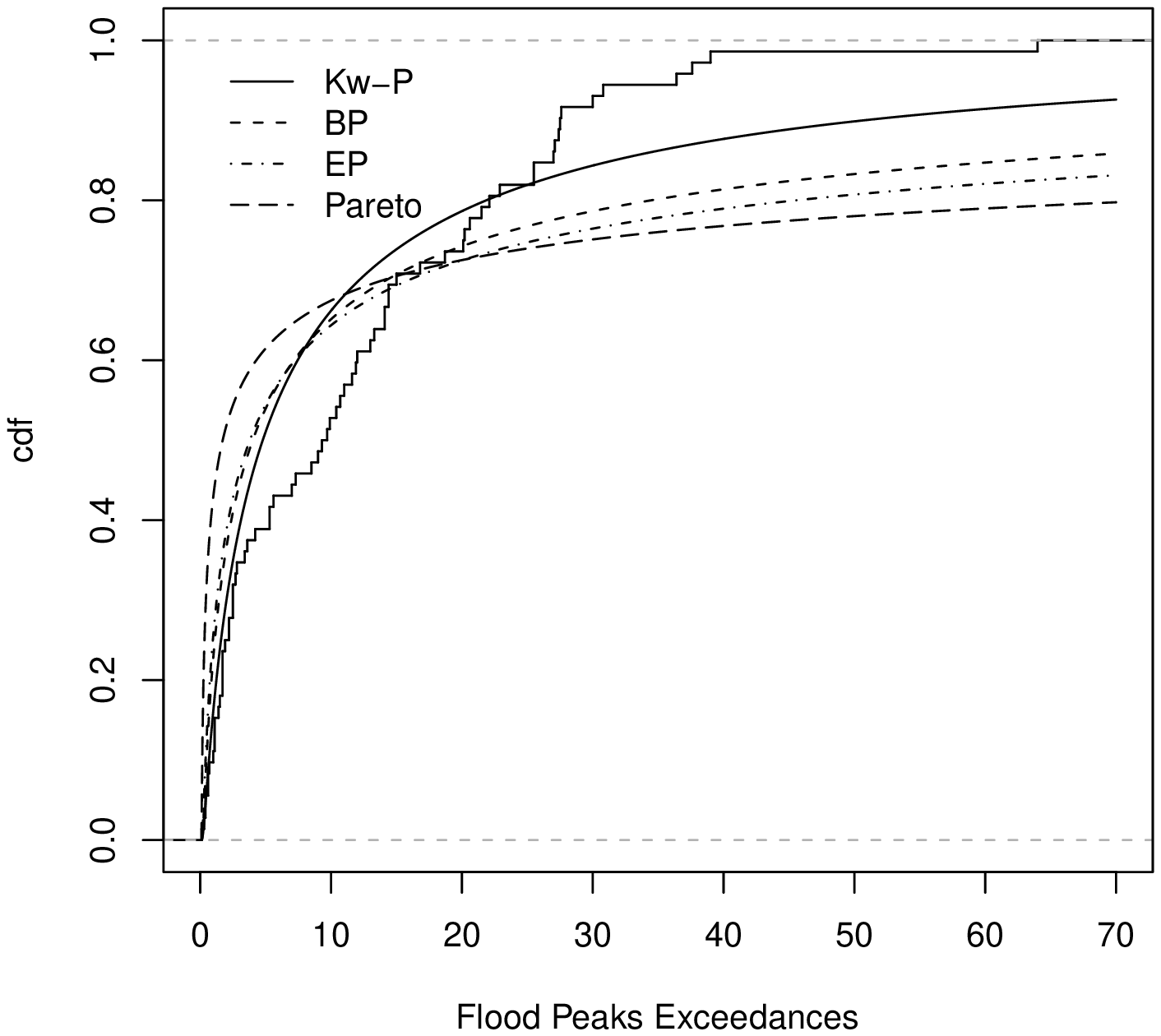}}
\caption{Estimated pdf and cdf from the fitted Kw-P, BP, EP and Pareto models for the exceedances of flood peaks data.}
 \label{figapli1}
\end{figure}

Plots of the estimated pdf and cdf of the Kw-P, BP, EP and Pareto models fitted to these data are given in Figure \ref{figapli1}. They indicate that the Kw-P distribution is superior to the other distributions in terms of model fitting.

Table \ref{tab4} gives the values of the Kolmogorov-Smirnov (K-S) statistic and  of $-2\ell(\hat{\boldsymbol{\theta}})$. From these figures, we conclude that the Kw-P distribution provides a better fit to these data than the BP, EP and Pareto models.

\ssection{Concluding remarks}

The well-known two-parameter Pareto distribution is extended by introducing two extra shape parameters, thus defining the Kumaraswamy Pareto (Kw-P) distribution having a broader class of
hazard rate functions. This is achieved by taking (\ref{CDFEP}) as the baseline cumulative distribution of the generalized class of Kumaraswamy distributions defined by Cordeiro and de Castro~\cite{cordeirocastro2011}. A de\-tai\-led study on the ma\-the\-ma\-ti\-cal properties of the new distribution is presented. The new model includes as special sub-models the Pareto
and exponentiated Pareto (EP) distributions (Gupta {\it et al}.~\cite{guptaetal1998}). We obtain the moment generating function, ordinary moments, order statistics and their moments and Rényi entropy. The estimation of the model parameters is approached by maximum likelihood and the observed information matrix is derived. An application to a real data set shows that the fit of the new model is superior to the fits of its main sub-models. We hope that the proposed model may attract wider applications in statistics.

\begin{center}
{\bf Acknowledgments}
\end{center}
 We gratefully acknowledge the grants from CAPES and CNPq (Brazil).

\appendix
\section{Figures and Tables}

\begin{table}[!htb]
{\footnotesize\begin{center}\caption{Mean estimates and biases of the MLEs of
$\beta$, $k$, $a$ and $b$.}\label{tabsimu} \vspace{0.2cm}
\begin{tabular}{cccccccr}
\toprule
$n$&&Parameter&&Mean&&bias\\
\midrule
30   &&$\beta$         &&0.0002    &&\ \ 1.4998\\
     &&$k$             &&1.6956    &&$-$0.6956\\
     &&$a$             &&4.9281    &&$-$4.4281\\
     &&$b$             &&0.8979    &&\ \ 1.6021\\
        \midrule
50   &&$\beta$         &&1.5002    &&$-$0.0002\\
     &&$k$             &&0.7398    &&\ \ 0.2602\\
     &&$a$             &&0.6686    &&$-$0.1686\\
     &&$b$             &&2.3205    &&\ \ 0.1795\\
        \midrule
100  &&$\beta$         &&1.5001    &&$-$0.0001\\
     &&$k$             &&0.9063    &&\ \ 0.0937\\
     &&$a$             &&0.6923    &&$-$0.1923\\
     &&$b$             &&2.4386    &&\ \ 0.0614\\
 \bottomrule
\end{tabular}
\end{center}}
\end{table}

\begin{table}[!htbp]
{\footnotesize\begin{center}\caption{Exceedances of Wheaton River flood data}
\label{tab1} \vspace{0.2cm}
\begin{tabular}{rrrrrrrrrrrr}
\toprule
1.7& 2.2& 14.4& 1.1& 0.4& 20.6& 5.3& 0.7& 1.9& 13.0&12.0 &9.3\\
1.4& 18.7& 8.5& 25.5& 11.6& 14.1& 22.1& 1.1& 2.5& 14.4& 1.7& 37.6\\
0.6& 2.2& 39.0& 0.3& 15.0& 11.0& 7.3& 22.9& 1.7& 0.1& 1.1& 0.6\\
9.0& 1.7& 7.0& 20.1& 0.4& 2.8& 14.1& 9.9& 10.4& 10.7& 30.0& 3.6\\
5.6& 30.8& 13.3& 4.2& 25.5& 3.4& 11.9& 21.5& 27.6& 36.4& 2.7& 64.0\\
1.5& 2.5& 27.4& 1.0& 27.1& 20.2& 16.8& 5.3& 9.7& 27.5& 2.5& 27.0\\
\bottomrule
\end{tabular}
\end{center}}
\end{table}

\begin{table}[!htb]
{\footnotesize\begin{center}\caption{Descriptive statistics}\label{tab2} \vspace{0.2cm}
\begin{tabular}{ccccccc}
\toprule
Min.  & $Q_{1}$   &  $Q_{2}$&    Mean &  $Q_{3}$  &Max.   &Var.\\
0.100 &   2.125 &  9.500&   12.200&  20.120 &64.000 &151.221\\
\bottomrule
\end{tabular}
\end{center}}
\end{table}

\begin{table}[!htbp]
{\footnotesize\begin{center} \caption{MLEs of the model parameters, the corresponding SEs (given in parentheses) and the statistics AIC, BIC and CAIC}\label{tab3} \vspace{0.2cm}
\begin{tabular}{lcccccccc}
\toprule
\multicolumn{5}{c}{Estimates}&&\multicolumn{3}{c}{Statistic}\\
\cmidrule{2-5} \cmidrule{7-9}
Model   &$a$      &$b$       &$k$       &$\beta$          & &AIC &BIC  & CAIC\\
\midrule
Kw-P    &2.8553   &85.8468   &0.0528    &0.1       & &548.4     &555.3 &548.8    \\
        &(0.3371) &(60.4213) &(0.0185)  &$-$       & &          &      &    \\
BP      &3.1473   &85.7508   &0.0088    &0.1       & &573.4     &580.3 &573.8    \\
        &(0.4993) &(0.0001)  &(0.0015)    &$-$       & &          &      &     \\
EP      &2.8797   &  1       &0.4241    &0.1       & &578.6     &583.2 &578.8    \\
        &(0.4911) & $-$      &(0.0463)  &$-$       & &          &      &     \\
Pareto  & 1       &  1       &0.2438    &0.1       & &608.2     &610.4 &608.2    \\
        & $-$     & $-$      &(0.0287)  &$-$       & &          &      &     \\
\bottomrule
\end{tabular}
\end{center}}
\end{table}

\begin{table}[!htbp]
{\footnotesize \begin{center}\caption{K-S and $-2\ell(\hat{\boldsymbol{\theta}})$  statistics for the exceedances of flood peaks data}\label{tab4} \vspace{0.2cm}
\begin{tabular}{lcccc}
\toprule
Model&Kw-P&BP&EP&Pareto\\
\midrule
K--S&0.1700&0.1747&0.1987&0.3324\\
$-2\ell(\widehat{\theta})$&542.4&567.4&574.6&606.2\\
\bottomrule
\end{tabular}
\end{center}}
\end{table}

\end{document}